\documentclass[fleqn,usenatbib]{mnras}

\usepackage{newtxtext,newtxmath}

\usepackage[T1]{fontenc}

\DeclareRobustCommand{\VAN}[3]{#2}
\let\VANthebibliography\thebibliography
\def\thebibliography{\DeclareRobustCommand{\VAN}[3]{##3}\VANthebibliography}

\usepackage{graphicx}	
\usepackage{amsmath}	
\usepackage{booktabs}
\usepackage{threeparttable}

\newcommand{\fer}{{\it Fermi}}
\newcommand{\mx}{{\it MAXI}}
\newcommand{\sw}{{\it Swift}}

\newcommand{\hzs}{{\rm\,Hz\,s^{-1}}}

\newcommand{\hx}{Her X-1}

\title[ELS of \hx]{The fading superorbital clock of Hercules X-1 in its 2025--2026 extended low state}

\author[Z. Liao et al.]{
Zhenxuan Liao$^{1}$\thanks{E-mail: liaozhenxuan@fjsmu.edu.cn},
Xinrui Tie$^{1}$,
Yufeng Chi$^{1}$,
Guanjun Lin$^{1}$
\\
$^{1}$School of Information Engineering, Sanming University, Jingdong Road 25, Sanming 365004, Fujian Province, People's Republic of China\\
}

\date{Accepted XXX. Received YYY; in original form ZZZ}

\pubyear{2026}

\begin{document}
\label{firstpage}
\pagerange{\pageref{firstpage}--\pageref{lastpage}}
\maketitle


\begin{abstract}
	We present two decades of X-ray monitoring of the eclipsing accreting
	pulsar Hercules X-1 with \textit{Fermi}/GBM, \textit{Swift}/BAT and
	\textit{MAXI}/GSC, centred on the extended low state (ELS) that began
	in 2025 May (the first such event in 22 years) and is still
	ongoing at the time of writing, $\sim$480~d after its onset. The 35-d
	superorbital clock does not simply stop in the ELS; it fades.
	The surviving modulation stays phase-locked to the pre-ELS clock
	until the flux of the main high state (the bright phase of the
	35-d cycle) has decayed to a few per cent of its normal level. The
	in-phase coherence is then lost, around 2026 January--March.
	We classify the pre-ELS
	pulse-frequency history into three torque states: spin-up, spin-down
	and flat. The spin derivative correlates positively with the
	main-high hard X-ray flux (Pearson $r\simeq0.7$), and the ELS hosts
	the longest sustained spin-down of the record,
	$\dot{\nu}\approx-4.7\times10^{-13}$~Hz~s$^{-1}$ over the
	$\gtrsim$380~d of \fer/GBM coverage. The 8--50 keV pulse-profile
	morphology is statistically indistinguishable across the torque
	states at fixed superorbital phase, and the shape parameters show no
	secular drift over the 18-yr record.
	These results show that the precessing warped accretion disc keeps
	its rhythm while the pulsar fades from direct view: the clock keeps
	its phase, the torque keeps tracking the accretion flux, and the beam
	keeps its shape. No free precession of the neutron star is required
	by the current monitoring data.
\end{abstract}




\section{Introduction}
\label{sec:intro}

Accreting X-ray pulsars are strongly magnetized neutron stars (NSs) in
binary systems, where material captured from a companion star is funnelled on to the
magnetic polar caps. As the magnetic and rotation axes are misaligned,
the source appears as a coherently pulsating X-ray source
\citep[for reviews, see][]{Caballero2012,Mushtukov2024}.
In most high-mass X-ray binaries, the
NS captures the fast stellar wind of an OB supergiant, or
material from the decretion disc of a rapidly rotating Be star
\citep[for reviews, see][]{Reig2011,Kretschmar2019,Weng2024}.
In low-mass systems the companion fills its Roche lobe,
and the transferred stream settles into
a persistent accretion disc before reaching the magnetosphere
\citep[for a review, see][]{Bahramian2024}.
The accreting plasma carries angular momentum, so the spin of the
NS evolves under the accretion torque, and the
spin-frequency history, measurable on time-scales from days to
decades, probes both the accretion mechanism and the distribution of
matter around the pulsar \citep{Bildsten1997,Malacaria2020}.

In the classical disc-torque model of \citet{Ghosh1979}, the disc is
truncated near the magnetospheric radius, where the magnetic stresses
of the dipole field overcome the ram pressure of the disc flow, and
the spin-up (SU) rate scales with luminosity as $\dot{\nu}\propto L^{6/7}$,
modulated by a fastness function $n(\omega_*)$ that vanishes at the
equilibrium spin period. In practice, however, the measured
$\dot{\nu}$--luminosity relation deviates from the canonical $6/7$
index and shows source-to-source diversity \citep{Sugizaki2017}: some
persistently accreting pulsars spin down steadily for years
\citep{Chakrabarty1997a}, and others reverse their torque abruptly with
little change in luminosity \citep{Chakrabarty1997b,CameroArranz2010}.
Such behaviour has been ascribed to flips between prograde and
retrograde discs \citep{Nelson1997,vanKerkwijk1998,OLeary2026}, to
transitions of the inner-disc state \citep{YiVishniac1999}, or to
excursions back and forth across the spin-equilibrium point under
modest changes of the accretion rate \citep{Ertan2021,Gencali2022}.
Beyond the accretion torque, accreting pulsars
show further periodicities that are tied neither to the pulsar spin
nor to the binary orbit.
In disc-fed systems the disc can become warped and precess
\citep{Pringle1996,Wijers1999,Ogilvie2001}, a leading candidate for
the origin of the superorbital flux cycles of tens to a few hundred
days observed in several X-ray binaries \citep[e.g.][]{Wojdowski1998,Chou2026}.
The NS itself has also been argued to precess freely,
leaving imprints on the pulse timing and profiles \citep{Brecher1972,Truemper1986}.
Nor does the flow remain regular on longer time-scales: superimposed
on the regular cycles, some X-ray binary systems occasionally sink into extended
(anomalous) low states, months-long episodes in which the bright
states largely collapse \citep[e.g.][]{Torpin2017,Chou2026}.
Which component carries the superorbital clock, and how the torque
responds when the flux collapses, remain open questions.
The continuous records of the \fer\ Gamma-ray Burst Monitor (GBM),
the \sw\ Burst Alert Telescope (BAT) and the Gas Slit Camera (GSC) of
the Monitor of All-sky X-ray Image (\mx) now allow these questions
to be tested directly in \hx, the classical low-mass pulsar binary, sometimes
classified as a rare intermediate-mass X-ray binary.

Discovered with \textit{UHURU} \citep{Tananbaum1972}, \hx\ showed
1.24-s pulsations and X-ray eclipses recurring on the 1.7-d binary
orbit \citep{Giacconi1973}.
The optical counterpart HZ Herculis is an evolved
$\sim2\,{\rm M}_{\odot}$ A-type star that fills its Roche lobe
\citep{Bahcall1972}. This companion mass
places \hx\ in the intermediate-mass X-ray binary class, of which it
has long been the only unambiguously identified member
\citep{Podsiadlowski2002}.
The source yielded the first cyclotron resonance scattering feature
detected in a cosmic X-ray source \citep{Truemper1978}, fixing its
magnetic field at a few $10^{12}$ G, and the line energy has itself
continued to evolve secularly ever since \citep{Staubert2019,Staubert2020}.
The \textit{UHURU} light curve revealed the $\sim$35-d on-off cycle with
its main-on and short-on states \citep{Giacconi1973,Jones1976},
whose carrier has been debated ever since: occultation by a tilted,
warped, precessing accretion disc
\citep{Katz1973,Petterson1975,LeahyFrost2025} versus free precession
of the NS itself
\citep{Staubert2009,Kolesnikov2022,Heyl2024,Kosec2026}, while
irregular absorption dips reveal further structure in the flow
\citep{Klochkov2008}.
On top of the regular cycle, \hx\ has repeatedly
fallen into extended low states (ELSs), in 1983 \citep{Parmar1985}, 1993
\citep{Vrtilek1994}, 1999--2000 \citep{Coburn2000} and 2003--2004
\citep{StillBoyd2004}. During each ELS, the 35-d flux modulation was
reported to vanish. A new such event, the first in 22 years, began
in 2025 May \citep{Mihara2026} and continues to the time of writing.

In this work we exploit two decades of continuous monitoring of
\hx\ by \fer/GBM, \sw/BAT and \mx/GSC to characterize
the behaviour of the 35-d clock, the spin torque and the pulse-profile
morphology through an ELS.
Section~\ref{sec:obs} describes the three monitoring records, and defines the torque states and the superorbital light-curve states used in this work.
Section~\ref{sec:results} presents our results: the
detection of a residual superorbital periodicity inside the ELS
(Section~\ref{sec:els_ls}), the phase-locking of the folded superorbital
profiles to the pre-ELS clock (Section~\ref{sec:els_profile}), the
torque--flux correlation across various torque states,
including the longest sustained spin-down (SD) of the record
(Section~\ref{sec:nudot_flux}), the invariance of the
pulse-profile shape across states at fixed superorbital phase,
and the absence of any secular drift of
the profile shape over the 18-yr record (Section~\ref{sec:profiles}).
Section~\ref{sec:disc} confronts these findings with accretion-torque
theory and with the disc- versus NS-precession debate, and
Section~\ref{sec:concl} summarizes.

\section{Observations}
\label{sec:obs}

\begin{figure*}
	\includegraphics[width=\textwidth]{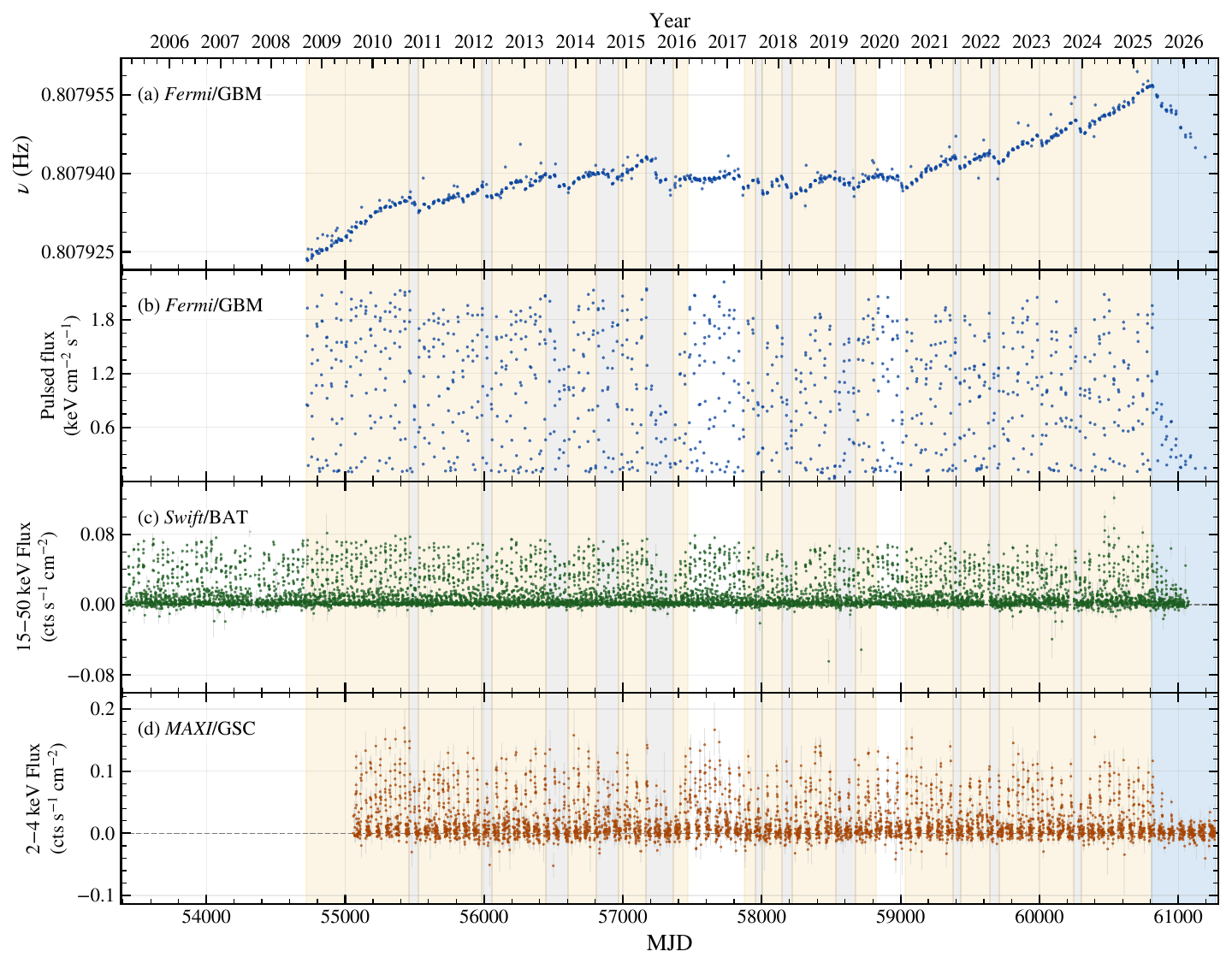}
	\caption{Two decades of monitoring of \hx.
		(a) Pulse-frequency history from \fer/GBM (12--50 keV).
		(b) Corresponding root-mean-square pulsed flux from
		\fer/GBM.
		(c) Daily 15--50 keV count rate from the \sw/BAT transient monitor, ending at MJD~61068.
		(d) Daily 2--4 keV count rate from \mx/GSC.
		The shaded bands mark the torque states defined in
		Section~\ref{sec:obs}: orange for SU episodes, grey for SD episodes, and blue for the 2025--2026 ELS, which started at the turn-on of superorbital cycle 212 (MJD~60805.3; cycle numbering of \citealt{LeahyWang2020});
		intervals of the GBM record left unshaded are ``Flat'' episodes, in which the local spin derivative is
		consistent with zero.
		The top axis gives calendar years.}
	\label{fig:overview}
\end{figure*}

Our study rests on three long-term monitoring records of
\hx: pulse-timing measurements from \fer/GBM, hard X-ray fluxes from
\sw/BAT, and soft X-ray fluxes from \mx/GSC. The three records span the past two decades (Fig.~\ref{fig:overview}).

\fer/GBM \citep{Meegan2009} views the entire sky that is not occulted
by the Earth with its 12 sodium-iodide detectors, allowing bright
accreting pulsars to be tracked on a daily basis.
Within the GBM Accreting Pulsar Program \citep{Malacaria2020}, the
1.24-s pulsations of \hx\ are phase-connected in the 12--50 keV band.
The pulsar-history file is publicly
available\footnote{\url{https://gammaray.nsstc.nasa.gov/gbm/science/pulsars/lightcurves/herx1.html}}
and lists, for every interval with a significant detection, the
barycentred and binary-corrected pulse frequency and its derivative,
together with the root-mean-square pulsed flux obtained from a
harmonic decomposition of the folded pulse profiles.

The BAT \citep{Krimm2013} on board \sw\ is a coded-mask hard X-ray
instrument whose transient monitor has followed \hx\ in the 15--50
keV band since 2005 February. We retrieved the public light
curves in the daily and per-satellite-orbit binnings\footnote{\url{https://swift.gsfc.nasa.gov/results/transients/HerX-1/}}.
The BAT record ends on 2026
January 28 (MJD~61068). The subsequent gap is instrumental: on 2026
February 11 the \sw\ mission suspended most science operations and
adopted a minimum-drag attitude to slow its orbital decay ahead of a
planned reboost mission \citep{Cenko2026}.

The \mx\ mission \citep{Matsuoka2009} on board the International
Space Station scans nearly the whole sky once per orbit with the GSC
\citep{Mihara2011}, which is sensitive in the 2--20 keV band. We use
the public \mx/GSC light curves of
\hx\footnote{\url{http://maxi.riken.jp/star_data/J1657+353/J1657+353.html}},
available in the daily and per-orbit binnings and in the full 2--20
keV band as well as the three standard sub-bands (2--4, 4--10 and
10--20 keV). Since the space-station orbit
precesses with a period of about 70 d and several GSC cameras are no
longer operational, the \mx\ record contains recurring gaps of up to
a few weeks. During the 2025--2026 ELS, moreover, the main-on states
of only every second superorbital cycle were sampled effectively
\citep{Mihara2026}.

Fig.~\ref{fig:overview} presents the three monitoring records together and marks
the states analysed in this paper. The pulse-frequency history
(panel a) shows the secular SU of \hx\ on which alternating
intervals of enhanced SU and of SD are superimposed
\citep{Deeter1981}. We divide the pre-ELS GBM record into torque
states with an objective, reproducible algorithm.
Points deviating by more than $5\sigma$
from a running
median of the frequency history are discarded; the cleaned history is
then interpolated on to a uniform grid, low-pass filtered with a
25-d Gaussian kernel to suppress the quasi-35-d torque sawtooth,
and split at the zero crossings of
the derivative of the smoothed curve. Episodes shorter than 50~d are
merged into their neighbours, and the boundaries are snapped to the
nearest significant local extremum of a lightly smoothed (10-d
kernel) version of the history. Each episode is classified by the
sign of a weighted linear fit to its frequencies: positive slope as
SU, negative as SD, and episodes with $|\dot{\nu}| < 5\times10^{-14}\hzs$
sustained for at least 120~d as ``Flat'' (shorter flat-looking transitions are
absorbed into their neighbours). This process yields 14 SU, 13 SD and 2 Flat
episodes, which tile the \fer/GBM record seamlessly up to the onset of the ELS
and are marked by the shaded bands in Fig.~\ref{fig:overview}; the
SU/SD boundaries agree closely with episodes identified by eye in
earlier work \citep{Klochkov2009}.
Panels (b)--(d) trace the
superorbital light-curve states: under normal conditions the 35-d
cycle repeats with a bright main-on (main high, MH) lasting about
seven binary orbits, a first low state (LS1), a fainter short-on
(short high, SH), and a second low state (LS2).
The new ELS that began in 2025 May,
at the turn-on of
superorbital cycle 212, MJD~60805.3, following the cycle numbering of
\citet{LeahyWang2020}, is shaded blue \citep{Mihara2026}. In what
follows we quantify how the light-curve states decay into this ELS,
whether a superorbital modulation persists within it, and how the
torque states evolve concurrently.

\begin{figure*}
\includegraphics[width=\textwidth, height=0.80\textheight, keepaspectratio]{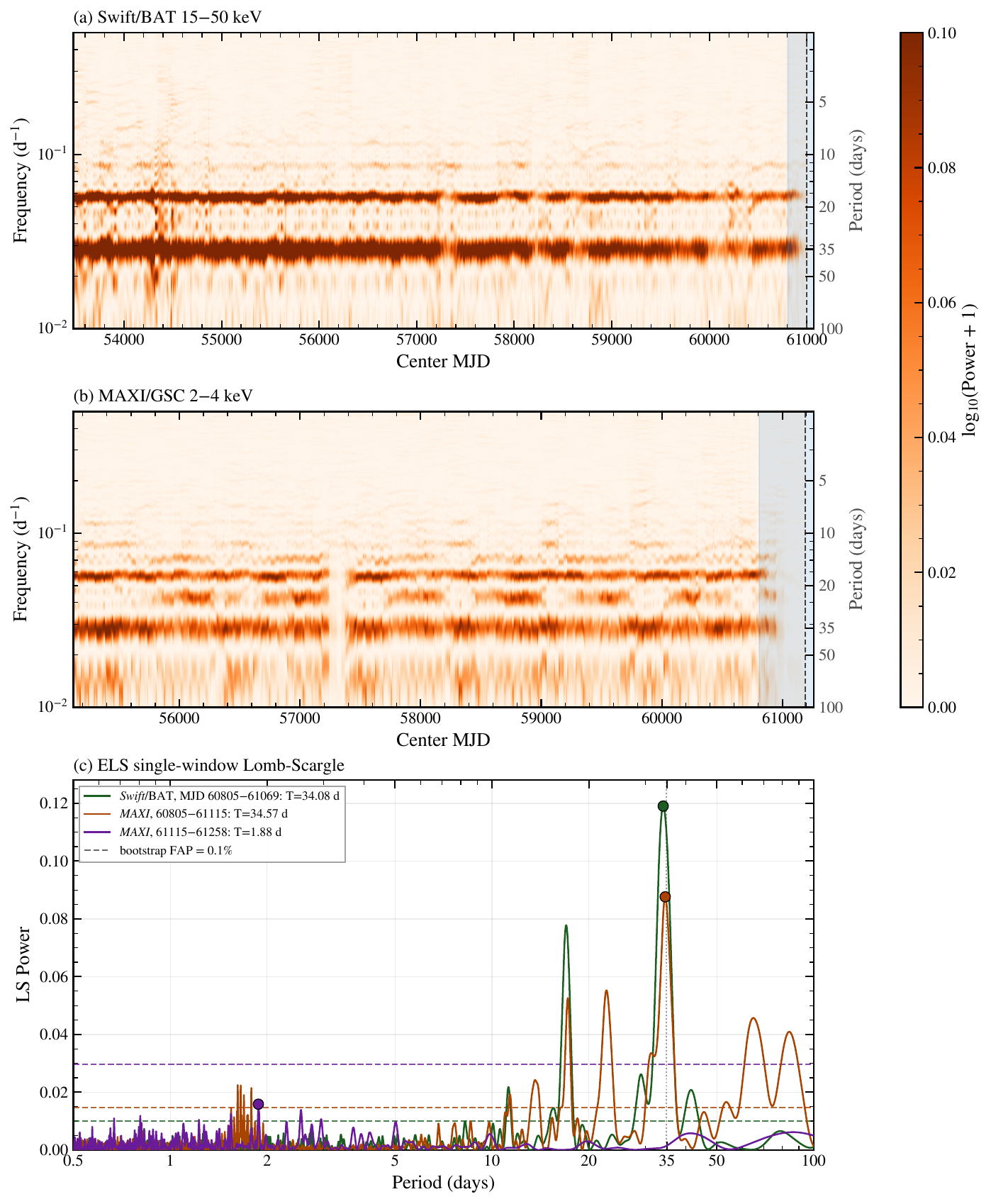}
\caption{The 35-d superorbital modulation of \hx\ through the
2025--2026 ELS. (a), (b) Running Lomb-Scargle periodograms (140-d
window, 10-d step) of the full \sw/BAT and \mx/GSC records; the blue
band marks the ELS.
Windows are plotted at their centre MJD; the dashed lines mark the
last window centres.
The colour scale is saturated at
$\log_{10}({\rm Power}+1) = 0.1$. (c) Single-window periodograms of
the ELS interval: \sw/BAT from the ELS start to the end of its record
(green, MJD 60805.3--61068.6, $N = 2777$) and \mx/GSC split at
MJD~61114.9 (the turn-on of superorbital cycle 221) into an early
part (brown, MJD 60805.3--61114.9, $N = 1949$) and a late part
(purple, MJD 61114.9--61258.2, $N = 945$). The divider is placed where
the running periodogram in panel (b) ceases to show sufficient power
near 35~d. Dashed horizontal lines mark the bootstrap global-FAP
threshold of 0.1 per cent for each data set.}
\label{fig:els_period}
\end{figure*}

\section{Results}
\label{sec:results}

\subsection{Superorbital period in the ELS}
\label{sec:els_ls}

We search for a superorbital periodicity in the 2025--2026 ELS with the
Lomb-Scargle periodogram \citep{Scargle1982,VanderPlas2018}, evaluated
with the \textsc{astropy} implementation \citep{Astropy2022} on the \sw/BAT and \mx/GSC orbit light curves.
Eclipsed orbital phases are excluded beforehand, keeping phases
0.10--0.90 for BAT and 0.14--0.86 for GSC (phase zero marks
mid-eclipse). No floating mean is fitted, since the monitor count rates
are already centred near zero in the low state. The frequency grid
spans 0.01--2~d$^{-1}$ with an oversampling factor of 100, and the
highest peak is refined by parabolic interpolation. Significance is
assessed with a permutation bootstrap: the fluxes are shuffled at fixed
timestamps and the maximum periodogram power over the whole grid is
recorded for each of $10^{4}$ realisations. Quantiles of this
maximum-power distribution provide global false-alarm probability (FAP)
thresholds that account for the look-elsewhere effect across the grid, and we show the 0.1 per cent level. To localise the signal in
time we also compute running periodograms in a 140-d window stepped by
10 d. The results are collected in Fig.~\ref{fig:els_period}.

The \sw/BAT record ends at MJD~61068.6 (Section~\ref{sec:obs}), so
its periodogram covers the ELS from its start to that date. The \mx\
record extends $\sim$200~d longer. We split it at MJD~61114.9, the
turn-on of superorbital cycle 221, into an early and a late part.
This division is motivated by the running periodogram itself:
beyond $\sim$MJD~61100
the 35-d ridge no longer shows sufficient power
(Fig.~\ref{fig:els_period}b).
The dominant peak of the \sw/BAT periodogram is $T = 34.08 \pm
0.03$~d, 12 times its 0.1 per cent global FAP threshold; the first
harmonic at $\sim$17.5~d is also clearly detected, as expected for
the strongly non-sinusoidal 35-d waveform. The early \mx/GSC part gives
$T = 34.57 \pm 0.04$~d (6 times its 0.1 per cent threshold), whereas
in the late \mx/GSC part no frequency passes its threshold at all: the
highest peak reaches only half of it.
The running periodograms
(Fig.~\ref{fig:els_period}a,b) show how the 35-d ridge, which
dominates the two-decade record, continues into the ELS at a reduced
level and then fades below detectability, without a measurable
frequency drift.
Given the known irregularity of the 35-d clock
\citep{Staubert2009} and the few cycles that each monitoring interval spans,
the $\sim$0.5~d difference between the \sw/BAT and early \mx/GSC
peaks is not unexpected. Because the period measured within the ELS
is sensitive to the chosen interval, in what follows we adopt the
cross-correlation cycle lengths of
Section~\ref{sec:els_profile} as the more robust measurement of the
superorbital clock.

A residual superorbital modulation persisting into an ELS has not been reported before.
In the four historical events (1983, 1993, 1999--2000 and
2003--2004) the 35-d flux modulation was found to vanish from the
monitor light curves. \textit{EXOSAT} measured only a weak source at low-state
level in 1983 \citep{Parmar1985}. The 1993 event was entered through a
mid-main-on collapse \citep{Vrtilek1994}. In dynamic power spectra the superorbital signal became undetectable within at
most three cycles around the 1999--2000 event \citep{Clarkson2003}, and
the 2003--2004 event left only per-cent-level residual flux
\citep{StillBoyd2004}. Yet the underlying clock kept
running in every case: the optical 35-d morphology was unchanged and
the pulse-profile evolution continued through the anomalous states
\citep{Staubert2009}. These non-detections, however, were
at least partly a matter of sensitivity: during an ELS the source sits at its low-state flux level,
which was sampled only sparsely and with limited exposure
by the monitors of those eras.
The deeper and more continuous
BAT and GSC records now show the behaviour of the residual
modulation directly: it remains clearly detectable in the early ELS
and then fades below the detection threshold as the state deepens
(Fig.~\ref{fig:els_period}). How far this residual modulation follows
the pre-ELS clock is addressed by the folded superorbital
profiles of Section~\ref{sec:els_profile}.

\subsection{The superorbital profile through the ELS}
\label{sec:els_profile}

A periodogram alone cannot say whether the residual ELS modulation
still follows the same clock as before: whether its phase
matches the pre-ELS cycle. We therefore fold the monitoring light
curves into average superorbital profiles, following the
cross-correlation (CC) timing method of
\citet{LeahyWang2020,LeahyWang2021}.

We use the orbit-binned \sw/BAT (15--50~keV) and \mx/GSC (2--20~keV)
records, exclude binary-eclipse phases beforehand (keeping orbital
phases 0.16--0.84, which also removes the pre-eclipse dips that would
otherwise alias the 1.7-d period into the CC), and measure the
weighted CC between the data and a 100-bin template over windows
$[t_n - 0.3\,P_{35}, t_n + 0.7\,P_{35}]$,
\begin{equation}
	\mathrm{CC}(t_n) = \frac{\sum_k R_k T_k \omega_k}{\sum_k T_k
		\omega_k}, \qquad \omega_k = 1/\epsilon_k,
	\label{eq:cc}
\end{equation}
where $R_k \pm \epsilon_k$ is the measured count rate in bin $k$,
$T_k$ is the template value, $t_n$ is the trial time of cycle $n$,
and $P_{35}$ is the mean superorbital period. The maxima of the CC
curve give the MH peak time of each cycle, which defines superorbital
phase zero; the template is rebuilt from the measured peak times and
the procedure is iterated until the timings stabilise.
The four-state division of the superorbital cycle adopted in this
work is: MH, phases 0.865--1.25; LS1, 0.25--0.435; SH, 0.435--0.71;
and LS2, 0.71--0.865, with the turn-on and decline transitions of
\citet{LeahyWang2021} merged into the corresponding high states.

\begin{figure}
	\includegraphics[width=\columnwidth]{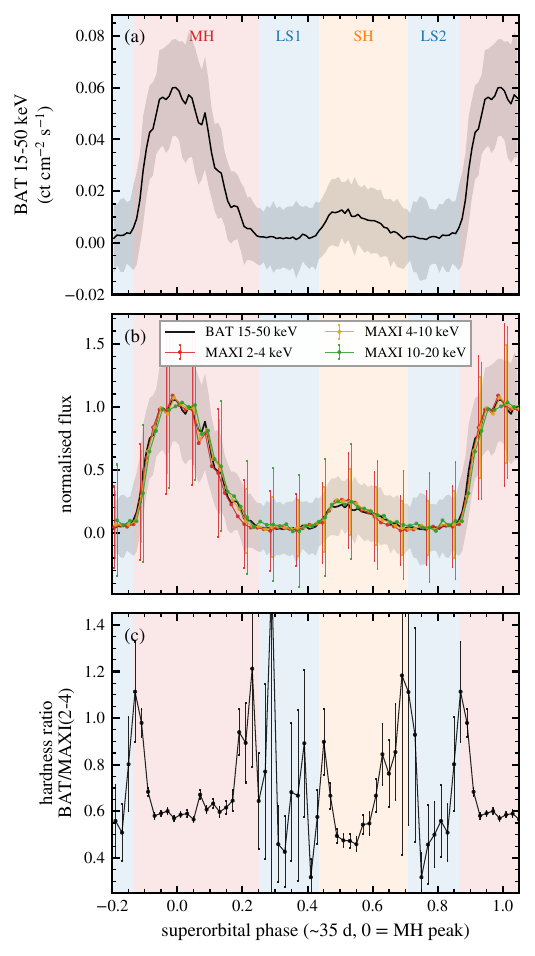}
	\caption{Average superorbital profiles of \hx\ before the ELS,
		constructed with the CC timing method described in
		Section~\ref{sec:els_profile}, with phase zero at the MH peak.
		Coloured bands mark the four-state division of the superorbital cycle
		defined in the same section (MH, LS1, SH and LS2); the same bands
		underlie all three panels. (a) \sw/BAT 15--50 keV profile from 212
		cycles (MJD 53416--60800), 100 phase bins; the grey band shows the
		$\pm1\sigma$ of the measurements within each phase bin,
		i.e.\ the cycle-to-cycle dispersion (the errors on the bin means are
		smaller by a factor $\sim\sqrt{n}$, where $n$ is the number of cycles
		contributing to the bin). (b) The \sw/BAT profile (black line and
		grey band) compared with the \mx/GSC 2--4, 4--10 and 10--20 keV
		profiles (50 phase bins), each normalised to its MH mean; error bars
		show the $\pm1$ standard deviation for every fourth bin and are
		offset horizontally between bands for clarity. The SH is relatively
		brighter at softer energies, and the persistent absorption dips near
		phases 0.02 and 0.06 are deeper in the softer bands. (c) Hardness
		ratio between the \sw/BAT 15--50 keV and \mx/GSC 2--4 keV bands; it
		hardens at both turn-ons and is meaningful only during the MH and SH
		states.}
	\label{fig:suporb_profile}
\end{figure}

\begin{figure}
	\includegraphics[width=\columnwidth]{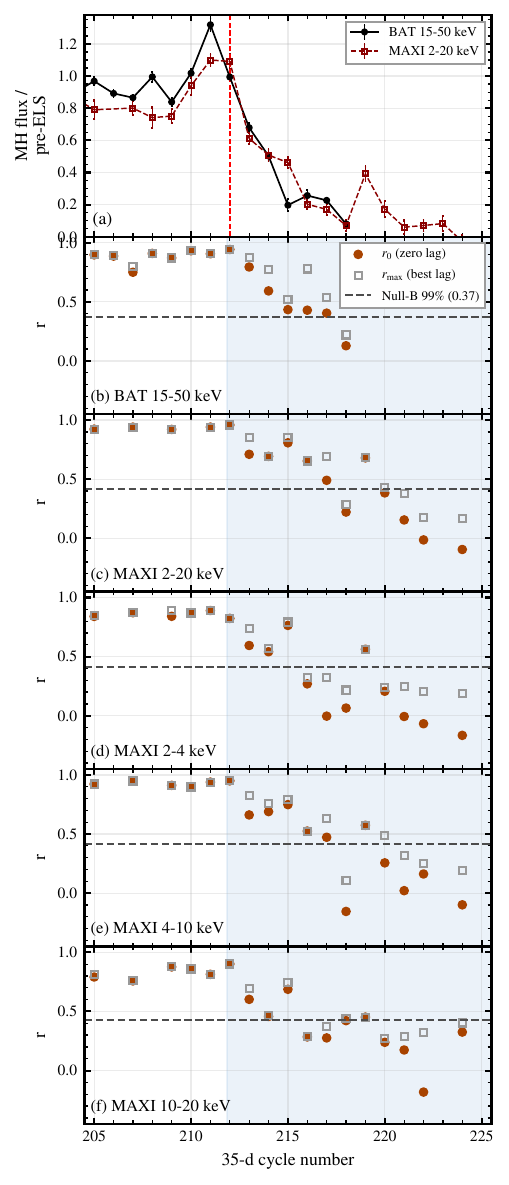}
	\caption{The fading 35-d clock of \hx\ through the 2025--2026 ELS.
		(a) MH flux per superorbital cycle (phase window 0.945--1.055) in
		units of the pre-ELS mean for \sw/BAT (black circles) and \mx/GSC
		2--20 keV (red squares), both measured in this work. (b)--(f)
		Per-cycle Pearson correlation of the single-cycle folded profile with
		the pre-ELS template of Fig.~\ref{fig:suporb_profile}, at zero lag
		($r_0$, filled circles; open circles mark cycles without MH coverage)
		and at the best lag within $\pm3.5$~d ($r_{\rm max}$, open squares).
		Dashed lines mark the 99th percentile of the permuted-bin null
		distribution. The blue shading marks the ELS from the turn-on of
		cycle 212; the \sw/BAT record ends at MJD~61068.6, so its entries
		stop at cycle 218 (cycle 219 is only partially covered).}
	\label{fig:els_decay_rmatch}
\end{figure}

\begin{figure}
	\includegraphics[width=\columnwidth]{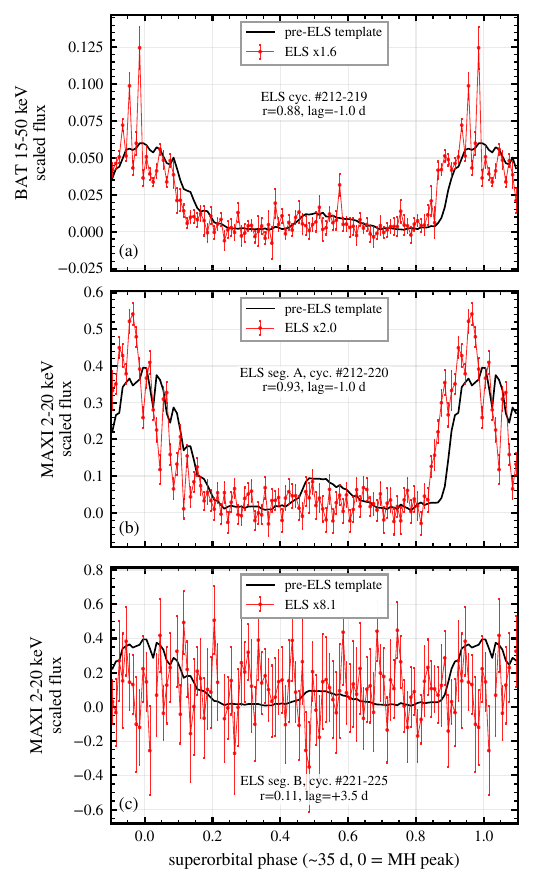}
	\caption{Folded ELS profiles (red) against the pre-ELS templates
		(black). The ELS profiles are scaled to the same phase-averaged mean
		as the template (factor given in each panel); the Pearson correlation
		$r$ and the lag of the residual MH bump are annotated. (a) \sw/BAT
		15--50 keV, all ELS cycles (\#212--219) combined (cycle 219 is only
		partially covered). (b) \mx/GSC 2--20 keV, ELS segment A (cycles
		212--220). (c) The same for segment B (cycles 221--225): no
		in-phase structure remains.}
	\label{fig:els_profile_match}
\end{figure}

\begin{figure*}
	\includegraphics[width=\textwidth]{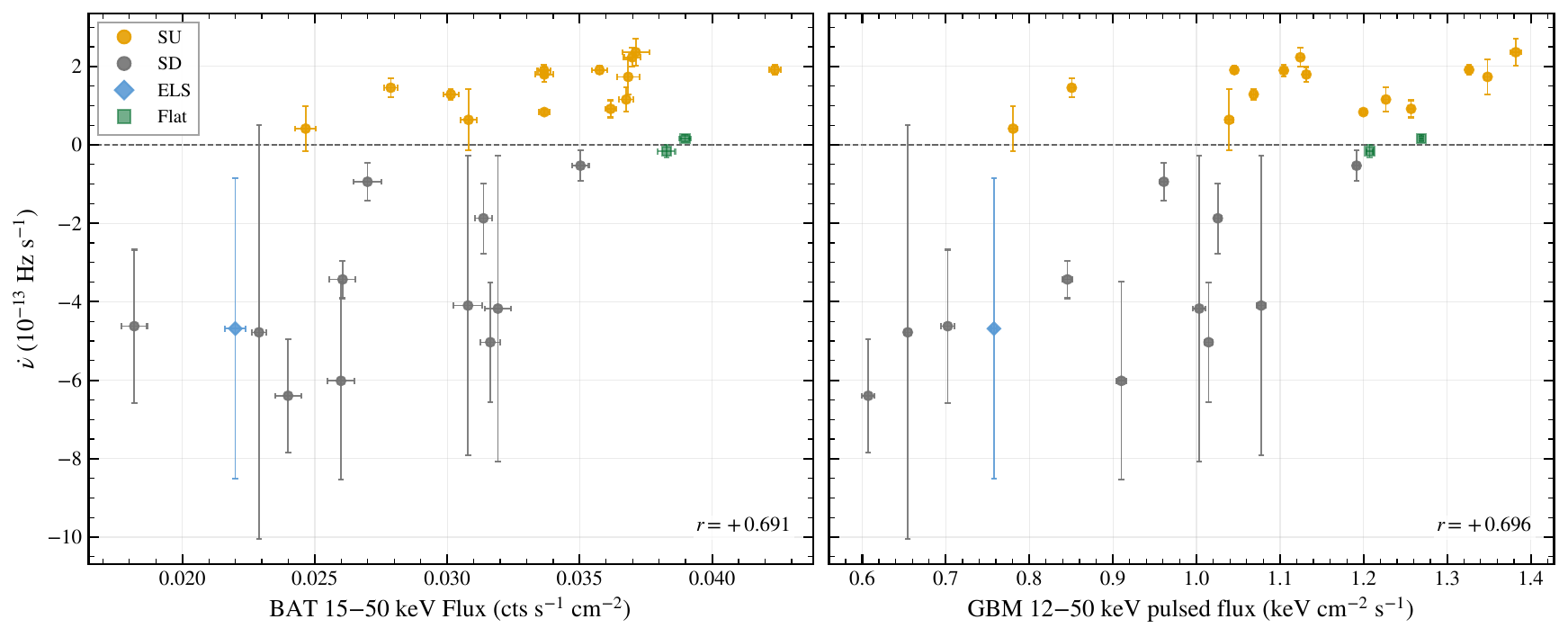}
	\caption{Spin derivative $\dot{\nu}$ versus MH hard X-ray flux for
		the episodes in Section~\ref{sec:obs}: (a) flux from
		the \sw/BAT 15--50 keV daily count rates; (b) root-mean-square pulsed
		flux from \fer/GBM (12--50 keV). In both panels only data within
		superorbital phases $-0.13$ to $+0.15$ of the MH peak enter. Orange
		circles mark SU intervals, grey circles SD, green squares Flat, and
		the blue diamond the 2025--2026 ELS. The flux error bars show the
		$1\sigma$ measurement uncertainties of the episode means; the
		$\dot{\nu}$ error bars are the $1\sigma$ uncertainties of a linear
		fit to the cycle-averaged frequencies, rescaled by
		$\sqrt{\chi^2_\nu}$ so that they include the cycle-to-cycle wander of
		the frequency level (episodes spanned by fewer than three 35-d
		cycles, and the ELS, use a point-level fit with the same rescaling).
		The Pearson $r$ annotated in the lower right of each panel includes
		all plotted points. A positive torque--flux correlation is visible in
		both measures.}
	\label{fig:nudot_flux}
\end{figure*}

Over the \sw/BAT record we obtain peak times for 218 cycles, 212 of
them before the ELS, with individual cycle lengths of $34.79 \pm
1.08$~d. The resulting average profiles
(Fig.~\ref{fig:suporb_profile}) reproduce the known morphology in
detail: the turn-on at phase $-0.12$ (flux first reaching 20 per
cent of the main-high peak), the persistent absorption dips at phases
0.98, 0.02 and 0.06 in the MH and at 0.55 and 0.62 in the SH, and
SH-to-MH flux ratios of 0.205 (\sw/BAT) and 0.244 (\mx/GSC), all in
agreement with \citet{LeahyWang2021} within the uncertainties.
Folding the \mx/GSC data on the \sw/BAT timings places the \mx/GSC
MH peak later by $0.6 \pm 1.3$~d, i.e.\ there is no measurable
energy-dependent offset of the peak. The hardness ratio between the
\sw/BAT and \mx/GSC 2--4 keV bands (Fig.~\ref{fig:suporb_profile}c)
hardens at both turn-ons, the signature of cold absorption by the
receding outer disc edge, to which the soft 2--4 keV denominator is
the most sensitive of the \mx/GSC bands. The ratio is meaningful only
during the MH and SH states: in LS1 and LS2 the soft flux drops
towards zero, leaving the ratio with large error bars and strong
point-to-point jitter.

To extend the profiles into the ELS, where the MH is too weak
for reliable CC timing, we extrapolate the peak times with a linear
ephemeris fitted to the last 25 pre-ELS cycles,
\begin{equation}
	t_{\rm peak}(N) = 60808.31 + 34.5644 \times (N - 212)\ \mathrm{MJD},
	\label{eq:ephemeris}
\end{equation}
i.e.\ a local cycle length of 34.564~d, noticeably shorter than the
two-decade mean of 34.79~d; such differences are expected, since the
superorbital period of \hx\ is known to wander on long time-scales,
typically shortening around the ELS
\citep{StillBoyd2004,Staubert2009,LeahyIgna2010}.
Figs~\ref{fig:els_decay_rmatch} and \ref{fig:els_profile_match} show
that through the ELS the 35-d clock does not simply stop: it fades.

Fig.~\ref{fig:els_decay_rmatch}a shows the MH flux per cycle in both
monitors: cycles 212--214 retain 50--100 per cent of the pre-ELS
level, cycles 215--218 fall to $\lesssim 25$ per cent, and beyond
cycle 219 only per-cent-level residuals remain, apart from a
transient partial recovery at cycle 219 in \mx/GSC.
Panels b--f show the per-cycle Pearson correlation coefficient $r$ of each
single-cycle folded profile with the pre-ELS template of
Fig.~\ref{fig:suporb_profile} (the null is built from 300
phase-permuted realisations per cycle): the in-phase match ($r_0$
above the 99 per cent permuted-bin null) survives through cycle 217
in \sw/BAT and through cycle 219 in \mx/GSC, and is lost afterwards.
The best-lag correlation $r_{\rm max}$ decays more slowly, i.e.\ some
residual structure persists for a few more cycles but no longer at
the extrapolated phase.

Fig.~\ref{fig:els_profile_match} shows the corresponding folded
profiles. We fold all \sw/BAT ELS cycles together (\#212--219, panel
a; cycle 219 is only partially covered), since its record ends at
MJD~61068.6: the residual MH bump remains at the expected phase
($r = 0.88$, lag $-1.0$~d). For \mx/GSC, whose record extends five
cycles longer, we fold separately the early and late parts of the
ELS, divided at the turn-on of cycle 221 (MJD~61114.9;
Section~\ref{sec:els_ls}): segment A (cycles 212--220) matches the
template at $r = 0.93$ with a lag of $-1.0$~d, while in segment B
(cycles 221--225) the match is gone ($r = 0.11$). The \mx/GSC
sub-bands behave identically within the noise: segment-A correlations
are $r = 0.86$, 0.91 and 0.86 for 2--4, 4--10 and 10--20~keV, and
none exceeds $r = 0.3$ in segment B.
The emerging picture is a fading 35-d clock: the residual modulation
remains locked to the pre-ELS phase for as long as it is visible, and
the in-phase coherence is lost only around cycles 219--221, once the
MH has faded to the per-cent level of its pre-ELS value and is no
longer discernible against the low-state floor. This is the behaviour
expected if the accretion-powered emission persists but is
continuously hidden from direct view, while the precessing disc keeps
its rhythm.

\subsection{The spin derivative versus X-ray flux across torque states}
\label{sec:nudot_flux}

We now correlate the spin derivative $\dot{\nu}$ with the
contemporaneous MH X-ray flux $\mathcal{F}$ across the
torque states. For every SU, SD, Flat and ELS episode of
Section~\ref{sec:obs} we measure $\dot{\nu}$ by a weighted linear
fit to the \fer/GBM spin frequency history. Since the remaining states
(SH, LS1 and LS2) dilute the flux dynamic range, the contemporaneous
flux is measured only in the MH state: following the
cross-correlation timing of Section~\ref{sec:els_profile}, a
measurement enters only if it falls within superorbital phases
$-0.13$ to $+0.15$ of the corresponding MH peak (218 good-time
intervals of $\sim$9.7~d each), i.e.\ the core of the MH state
defined in Section~\ref{sec:els_profile}. Within each cycle the
in-window \sw/BAT daily count rates are averaged weighted by the
partially coded exposure, and the \fer/GBM pulsed fluxes with
inverse-variance weighting.

The result is shown in Fig.~\ref{fig:nudot_flux} for the 28 episodes
with sufficient MH coverage. A positive correlation in the sense
expected for disc accretion \citep{Ghosh1979} is present for both
flux measures: Pearson $r = +0.69$ ($p = 5\times10^{-5}$) against the
\sw/BAT MH flux, and $r = +0.70$ ($p = 4\times10^{-5}$) against the
\fer/GBM MH pulsed flux, with the ELS point included.
The correlation is genuinely produced by the bright-state points:
repeating the test with full-cycle flux averages largely eliminates it
($r \simeq +0.07$ for \sw/BAT, $r \simeq +0.41$ for the GBM
pulsed flux), because the cycle-averaged flux of \hx\ varies by only
a factor of $\sim$2 over the whole record. 
The \fer/GBM value is less strongly diluted, because its significant
detections occur only when the
pulsations are bright enough to be measured, preferentially in the
MH, so that even its full-cycle average retains a bright-state
bias.
The two Flat episodes sit at $|\dot{\nu}| \lesssim
1\times10^{-13}\hzs$ at fluxes comparable to the SU episodes, as
required by their definition. The ELS reaches the low-flux end of the
correlation without departing from it, at $\dot{\nu} =
-(4.7\pm0.3)\times10^{-13}\hzs$ and with an MH flux among the lowest
of the sample, about 40 per cent below the normal SU level, although
the MH window list covers only the early ELS cycles 212--217.

Sustained from the turn-on of cycle 212 to the end of the GBM record
($\gtrsim$380~d of documented decline, and continuing), the ELS
SD is the longest SD episode in the \fer/GBM history of \hx,
although not the largest in magnitude (the 2017--2018 episodes
reached $\dot{\nu}\approx-6\times10^{-13}\hzs$ over $\sim$50--100~d)
and the first to coincide with the collapse of the MH states. We
return to its interpretation in Section~\ref{sec:disc_torque}.

We also carried out the point-wise analogue of this test, computing
three-point Lagrange derivatives of the frequency history and pairing
them with \sw/BAT fluxes interpolated to the derivative epochs by
inverse-distance weighting of the three nearest daily points. The
$\sim$500 resulting $\dot{\nu}$--flux pairs show no correlation
($r \simeq 0$), and a variance decomposition confirms that this is
not an artefact of measurement noise, which contributes only $\sim$3
per cent of the observed derivative variance: on few-day time-scales
the spin derivative of \hx\ is dominated by red (random-walk) torque
noise, and the flux--torque coupling emerges only after averaging
over the $\sim$100-d state episodes. This mirrors the ``wanderer''
behaviour established for \hx\ by \citet{Bildsten1997} and the
$\omega^{-2}$ timing-noise spectrum measured by \citet{Serim2023}.

\subsection{Pulse-profile morphology: state independence and long-term stability}
\label{sec:profiles}

Each entry of the \fer/GBM pulsar history integrates exactly one
eclipse-free binary orbit, from eclipse egress to ingress (orbital
phases 0.070--0.930, where phase zero marks mid-eclipse), so that
every measurement already averages the out-of-eclipse orbital phase
uniformly. Each entry provides the coefficients of a six-harmonic
Fourier decomposition of the folded pulse profile in three energy
bands (8--12, 12--25 and 25--50 keV), together
with the associated uncertainties and the number of pulses entering
each measurement \citep{Malacaria2020}. We characterize the profile shape with three
amplitude-based parameters: the harmonic purity (the fraction of
the combined power of the first six harmonics carried by the
fundamental), the pulse modulation index ${\rm PMI}=(F_{\rm
	peak}-F_{\rm min})/(2F_{\rm rms})$, and the full width at half
maximum (FWHM) of the profile, together with two shift-invariant harmonic
phase combinations, $\delta_2 = \angle H_2 - 2\angle H_1$ and
$\delta_3 = \angle H_3 - 3\angle H_1$, the analogues of the classical
Fourier phase parameters of variable light curves
\citep{SimonLee1981}, where $\angle H_k$ is the phase angle of the
$k$-th harmonic. The $\delta$ combinations are immune to any rigid
shift of the profile, a property that is essential here: the phase
zero of the published coefficients is arbitrary from entry to entry
(the measured fundamental phases are uniformly distributed),
so absolute pulse-arrival phases are not available from this data product, while
$\delta_2$ and $\delta_3$ retain the full phase-encoded shape
information.
We adopt the 12--25 keV band as the primary channel: both the pulsed
spectrum and the NaI effective area peak there, giving the highest
harmonic signal-to-noise per detection (median S/N $\simeq$53, versus
$\simeq$13 and $\simeq$22 in the adjacent bands).

Because the pulse profile is known to evolve systematically over the
35-d cycle \citep{Deeter1998,Scott2000,Staubert2009}, a comparison of different torque
states is meaningful only at fixed superorbital phase.
We therefore restrict every sample to the peak of the MH, using
detections within superorbital phases $-0.13$ to $+0.15$ of the MH
peak (the same MH windows as for Fig.~\ref{fig:nudot_flux}). The
four samples are taken from the torque state episodes of
Section~\ref{sec:obs}: SU (660 detections, 421\,306 pulses), SD (165
detections, 103\,544 pulses), Flat (106 detections, 70\,021 pulses)
and the ELS (24 detections, 15\,441 pulses).
For each sample we reconstructed the profile of every detection from
its harmonic coefficients, and aligned the profiles by a rigid phase
shift that places each main peak at the same reference phase.
The mean profile is then the pulse-count-weighted mean of the
aligned individual profiles, and their weighted standard deviation (the
dispersion among detections, bias-corrected for the unequal
pulse-count weights via the effective sample size $N_{\rm eff} =
(\sum_i w_i)^2/\sum_i w_i^2$) gives the error bars of the mean
profiles.

\begin{figure}
\includegraphics[width=\columnwidth]{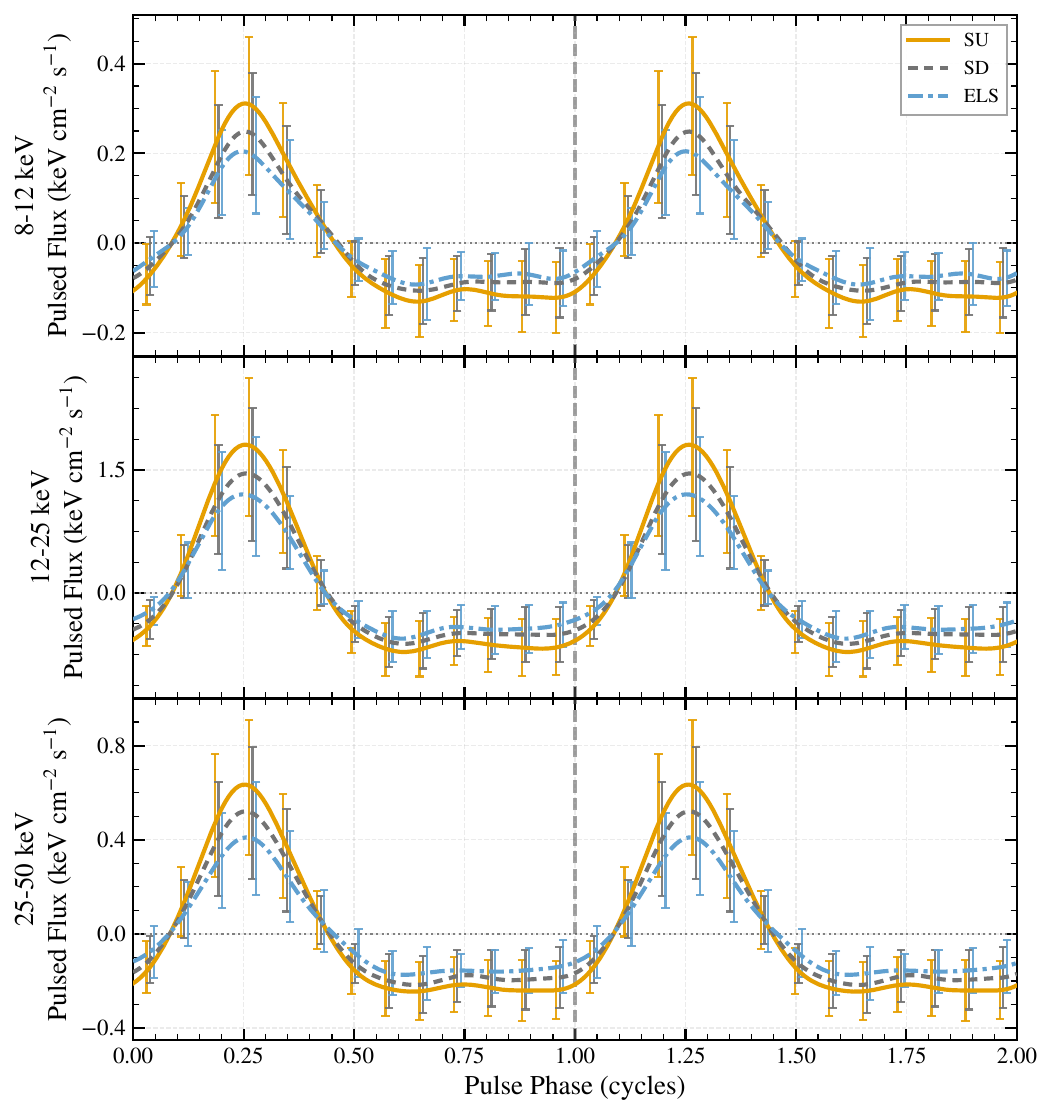}
\caption{Weighted mean pulse profiles of \hx\ at the peak of the MH
	(detections within superorbital phases $-0.13$ to $+0.15$ of the MH
	peak), in the three \fer/GBM energy bands. Three samples are compared: SU
	(amber solid; $N=660$), SD (grey dashed; $N=165$) and the ELS
	(sky-blue dot-dashed; $N=24$). Error bars mark the $\pm1\sigma$
	weighted dispersion among the individual profiles, evaluated at 13
	phase bins per cycle; two pulse cycles are shown. At fixed
	superorbital phase the profile morphology is statistically
	indistinguishable across the three samples; only the amplitude tracks
	the accretion state. The Flat-state profile ($N=106$, not shown) is
	consistent with SU to within $\sim$7 per cent in $F_{\rm rms}$.}
\label{fig:profiles}
\end{figure}

Fig.~\ref{fig:profiles} compares the weighted mean profiles of the
SU, SD and ELS samples and Table~\ref{tab:profile_stats} collects the
corresponding statistics. At this fixed superorbital phase, and in
all three bands, the profile is a single, fundamental-dominated peak
whose morphology is statistically indistinguishable in
the four samples (purity 76--82 per
cent, PMI 1.38--1.45, without state dependence), while $F_{\rm rms}$
scales with the accretion state.
Relative to SU, the mean $F_{\rm rms}$ is 81--83 per cent for SD and
66--68 per cent for the ELS, essentially band-independent. The Flat
sample matches SU within
$\sim$7 per cent and is therefore not shown separately in
Fig.~\ref{fig:profiles}. The band-independent amplitude ratios show
that the profile shape scales self-similarly with the accretion rate
while its gross morphology is unchanged.
Because GBM is a non-imaging collimated instrument, the history
measures only the pulsed component of the flux. The unpulsed baseline
is not available from this data product, so a classical pulsed
fraction cannot be constructed. The PMI is instead a purely
pulsed-domain shape index: it equals $\sqrt{2}$ for a purely
sinusoidal profile, and it is mildly biased low at low
S/N (the per-detection PMI anti-correlates with the
harmonic S/N at $r\simeq-0.7$), so the near-identical PMI across
torque states is a conservative statement of shape invariance.

\begin{figure}
\includegraphics[width=\columnwidth]{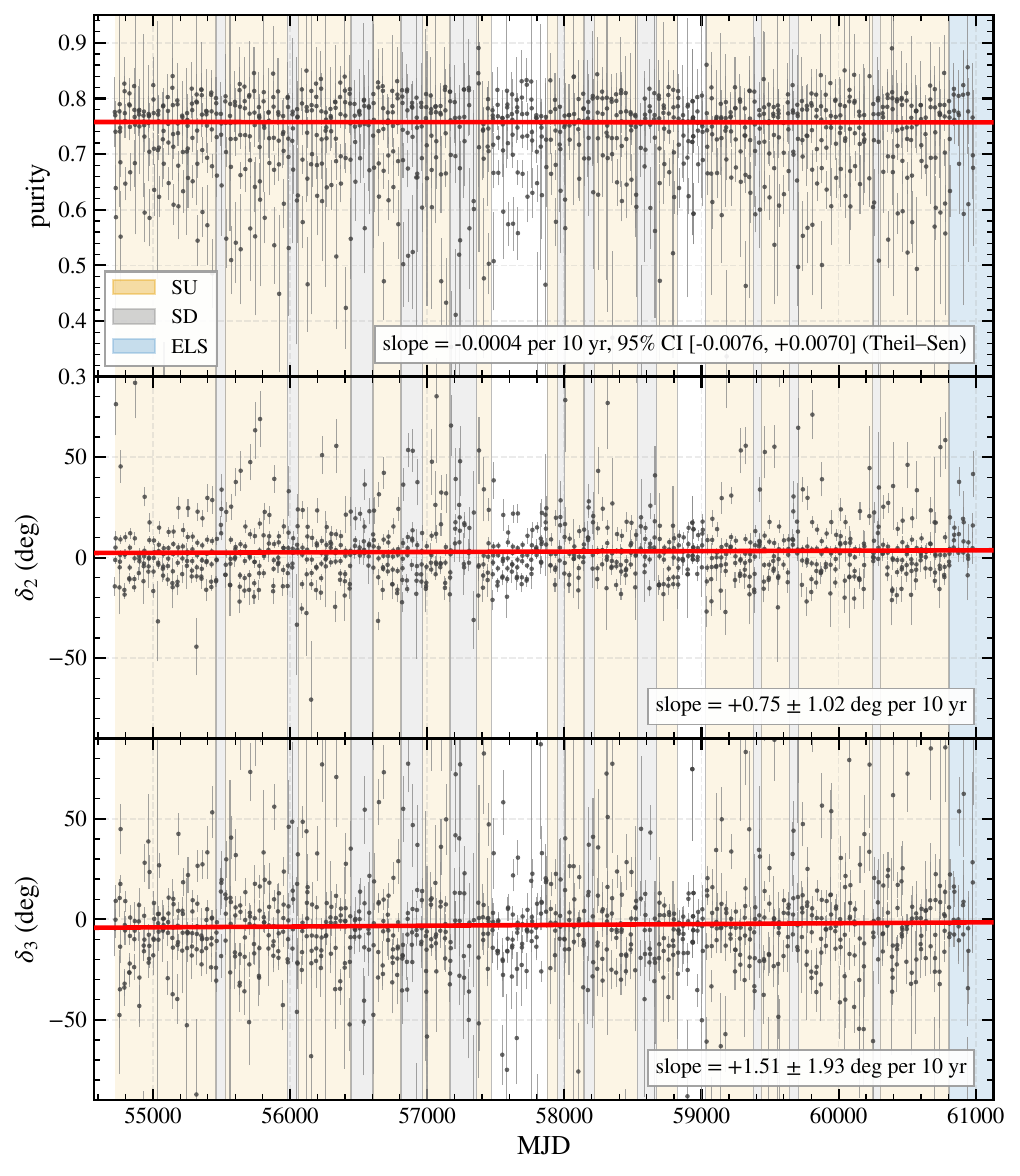}
\caption{Full-record stability of the profile shape of \hx\ in the
	primary 12--25 keV \fer/GBM band: harmonic purity (top) and the
	shift-invariant harmonic phases $\delta_2$ (middle) and $\delta_3$
	(bottom) for every MH detection, with the torque states shaded as in
	Fig.~\ref{fig:overview}. Red lines
	are linear fits (Theil--Sen for the purity, weighted least squares
	for the circular phases); the annotated slopes are all consistent
	with zero. }
\label{fig:stability}
\end{figure}

The invariance also holds in time.
For the time-resolved analysis we work per detection: the $\delta$
phases are computed directly from the coefficients of each entry of
the \fer/GBM pulsar history, without any alignment and free of the S/N bias of
the amplitude-based indices; they are therefore the cleanest
per-detection shape probes.
Fig.~\ref{fig:stability} shows the purity, $\delta_2$ and $\delta_3$
of the primary band for every MH detection across the 18-yr record,
with the torque states shaded as in Fig.~\ref{fig:overview}. Robust
(Theil--Sen) and weighted least-squares fits give drift rates
consistent with zero: the purity slope is $-0.0004$ per decade (95
per cent confidence interval $[-0.008, +0.007]$), and the harmonic
phases drift by less than a degree per decade ($\delta_2$:
$+0.68\pm0.55$; $\delta_3$: $+0.59\pm1.01$ degrees per decade). The
adjacent bands return the same null result at looser precision.
The point-to-point scatter in the figure exceeds the formal errors. It
is dominated by intrinsic cycle-to-cycle variations of the profile
shape and by the non-uniform sampling of superorbital phase within
the MH window, and it does not bias the slope estimates. We caution
that the amplitude-based estimators are mildly flux-dependent (the
PMI bias noted above).
One marginal feature deserves a note: at the MH peak the ELS points sit
$\sim+10\degr$ above the pre-ELS mean in $\delta_2$ (about $3\sigma$
in the primary band), but the offset is not confirmed in the 25--50
keV band, anti-correlates with the falling S/N, and
is sensitive to the choice of superorbital phase window. We therefore
do not claim it as a shape change and flag it for follow-up with
pointed observations.

The emerging picture is that of a stationary X-ray beam: at fixed
superorbital phase, the 8--50 keV pulse profile of \hx\ keeps the
same fundamental-dominated single-peak morphology across the SU, SD,
Flat and ELS states, and over the 18-yr record its shape parameters
are stationary to within a fraction of a per cent (amplitudes) and of
order a degree (harmonic phases) per decade; only the pulsed
amplitude tracks the accretion rate.

\begin{table}
	\centering
	\caption{Statistics of the weighted mean pulse profiles of
		Fig.~\ref{fig:profiles} (MH-peak phase only). $F_{\rm rms}$ is the
		mean of the per-detection root-mean-square pulsed fluxes; ${\rm
			PMI}=(F_{\rm peak}-F_{\rm min})/(2F_{\rm rms})$, which equals
		$\sqrt{2}$ for a purely sinusoidal profile; purity is the power
		fraction of the fundamental among the first six harmonics of the mean
		profile.}
	\label{tab:profile_stats}
	\begin{threeparttable}
		\begin{tabular}{llccc}
			\toprule
			State & Band & $F_{\rm rms}$ & PMI & Purity \\
			& (keV) & (keV\,cm$^{-2}$\,s$^{-1}$) & & (\%) \\
			\midrule
			SU\tnote{a} & 8--12 & 0.160 & 1.40 & 79.6 \\
			& 12--25 & 0.897 & 1.43 & 76.2 \\
			& 25--50 & 0.319 & 1.41 & 78.1 \\
			\midrule
			SD\tnote{b} & 8--12 & 0.133 & 1.40 & 79.8 \\
			& 12--25 & 0.726 & 1.45 & 76.5 \\
			& 25--50 & 0.258 & 1.44 & 78.6 \\
			\midrule
			ELS\tnote{c} & 8--12 & 0.107 & 1.39 & 80.8 \\
			& 12--25 & 0.606 & 1.43 & 78.6 \\
			& 25--50 & 0.210 & 1.38 & 81.6 \\
			\bottomrule
		\end{tabular}
		\begin{tablenotes}
			\item[a] 660 detections (421\,306 pulses), MH-peak
			in the SU torque state.
			\item[b] 165 detections (103\,544 pulses), MH-peak
			in the SD torque state.
			\item[c] 24 detections (15\,441 pulses), MH-peak
			in the ELS.
		\end{tablenotes}
	\end{threeparttable}
\end{table}

\section{Discussion}
\label{sec:disc}

\subsection{The accretion torque of \hx\ in context}
\label{sec:disc_torque}

The positive $\dot{\nu}$--$\mathcal{F}$ correlation of
Section~\ref{sec:nudot_flux} is the qualitative signature expected
when the spin evolution of a disc-fed pulsar is dominated by the
torque exerted by the accreted material: in the \citet{Ghosh1979}
picture the torque scales steeply with the accretion rate,
$\dot{\nu}\propto L^{6/7}$ modulated by the fastness function
$n(\omega_*)$.
Quantitatively, however, \hx\ shares the fate
of the other persistent disc-fed pulsars: the correlation is
significant but loose, with about an order of magnitude of scatter in
$\dot{\nu}$ at fixed flux (Fig.~\ref{fig:nudot_flux}).
The same pattern appears in the BATSE census of \citet{Bildsten1997},
in which transient outbursts
track the torque--luminosity relation closely while persistent sources
wander between SU and SD at nearly constant luminosity.
In the ensemble test of
\citet{Sugizaki2017}, the normalization of the \citet{Ghosh1979}
relation likewise deviates from the measurements by up to a factor
of a few.
For \hx, the earlier \sw/BAT analysis of \citet{Klochkov2009} found a
$\dot{P}$--flux correlation whose significance was driven mainly by a
few strong SD, low-flux points, superimposed on an otherwise
uncorrelated cloud near $\dot{P}=0$.
Our measurement improves on that early result in two respects: the
torque-state episodes are segmented objectively from the frequency
history itself rather than by eye, and the fluxes are restricted to
the MH state, removing the dilution by the remaining states (SH, LS1
and LS2). Even so, the residual scatter remains large.

Three effects plausibly contribute to that scatter. First, red torque
noise: the spin derivative of \hx\ fluctuates as an $\omega^{-2}$
process that dominates measurement noise on all time-scales longer
than $\sim$10~d \citep{Serim2023}, and our point-wise analysis
(Section~\ref{sec:nudot_flux}) shows this noise directly: individual
observations therefore carry stochastic $\dot{\nu}$ offsets unrelated to
the instantaneous flux, and only episode-averaged torques trace the
flux coupling. Secondly, \hx\ accretes close to the corotation radius,
where the torque is intrinsically bistable: the magnetospheric
boundary can switch between SU and SD configurations at
almost fixed accretion rate \citep{Lovelace1995}, an idea already
invoked by \citet{Klochkov2009} for the near-zero-torque cloud. The
two Flat episodes identified here are the natural
observational counterpart of that regime, and the interleaving of SU
and SD points at comparable fluxes in Fig.~\ref{fig:nudot_flux} is its
direct manifestation in our data.
Thirdly, the measured pulse frequency may also be modulated by free
precession of the NS \citep{Kolesnikov2022}, which would
blur the instantaneous $\dot{\nu}$--flux mapping at the level of
individual 35-d cycles (Section~\ref{sec:disc_clock}).

The ELS extends the correlation into the low-flux regime and ties
the torque behaviour to the light-curve phenomenology discussed in
this paper. That the collapse of the MH flux coincides with the
longest sustained SD of the record is qualitatively consistent with
disc-torque theory: the effective accretion rate on to the
magnetosphere falls when the inner-disc structure changes, and the
magnetospheric radius moves out towards corotation. A deeper drop
would push it beyond corotation into the propeller regime, where the
SD would be strongly enhanced \citep{LeahyIgna2010}. That the
measured rate stays within the range of historical SD episodes
indicates that this threshold was approached but not crossed.
Indeed, the necessity of the propeller mechanism for such transitions has
recently been questioned \citep{Tsygankov2026b}.

We note, however, that the torque responds far less than
proportionally to the observed flux: while the MH and pulsed fluxes
decay by an order of magnitude, the SD rate does not strengthen
accordingly. This is consistent with the historical interpretation of
anomalous low states as \emph{obscuration} rather than true
quiescence \citep{Parmar1985,Vrtilek1994,Still2001}: a substantial fraction of
the accretion luminosity is hidden from our line of sight by the
increasingly warped inner disc, so the measured flux underestimates
the $\dot{M}$ that sets the torque.
Two further considerations support this reading. First, the
near-total disappearance of the hard-band MH flux cannot be produced
by cold absorption alone, which leaks at these energies, but instead
requires the covering fraction of the screen to approach unity, while the
torque independently limits the true $\dot{M}$ drop to a moderate
value.
Secondly, a genuine gradual shutdown would drive the
magnetospheric radius beyond corotation and strengthen the SD
beyond the historical range, which is not seen. On the contrary,
pulsations at the extrapolated period and a reflection-dominated
spectrum persist into the deepest part of the ELS
\citep{Tsygankov2026}, so accretion on to the polar caps never
stopped.
Under this reading, the ELS is an
extreme point of the same $\dot{\nu}$--$\mathcal{F}$ relation that
governs the normal torque states, displaced downward in flux by
screening rather than shifted in torque physics, in line with the
phase-locked residual clock of Sections~\ref{sec:els_ls} and
\ref{sec:els_profile}.

A caveat of coverage: the \sw/BAT record ends at MJD~61068.6, while
the fading clock is still partly trackable there,
so the flux-side narrative of the late
ELS rests on \mx/GSC alone. The torque side is unaffected: the GBM
record extends to MJD~61190 with the SD rate stable within the
historical range, and the pulse period measured at MJD~61221--61227
\citep{Tsygankov2026} lies on the extrapolated trend.

\subsection{Which carries the 35-d clock? Accretion disc precession versus NS precession}
\label{sec:disc_clock}

The carrier of the 35-d superorbital periodicity of \hx\ has been debated for
decades. Occultation by a tilted, warped, precessing accretion disc
explains the flux cycle and its two high states
\citep{Katz1973,Petterson1975,LeahyFrost2025} and, via
partial occultation of the beam by the inner disc edge, the
systematic pulse-profile evolution with superorbital phase
\citep{Deeter1998,Scott2000}; alternatively, the NS itself
may precess freely, making the profile variations trace the star
while the flux traces the disc \citep{Staubert2009}.
A crucial constraint came from the same group: \citet{Staubert2013}
showed that the pulse-profile variations repeat cycle by cycle with
the same 35-d clock as the flux turn-ons, jitter included. A freely
precessing crust, driven by its own internal dynamics, would not be
expected to track the disc's irregular period so closely. This leaves
two possibilities: a single clock, with the disc driving both the
flux and the profiles, or a NS precession whose period itself
changes by up to $\sim$2.5 per cent every few years
\citep{Postnov2013}.
Profile complexity is not, by itself, an
argument for a second clock either: a static spin-disc misalignment
(known to be substantial in \hx\ \citep{Doroshenko2022,Heyl2024})
already produces asymmetric pulse profiles and spin-periodic
modulation of the accretion flow in simulations
\citep{Mushtukov2026}.

The debate has been revived from two directions. IXPE polarimetry
found the polarization angle to change by $9.2\degr\pm1.8\degr$
between the 2022 and 2023 MH states, and inferred a geometry of the
spin and magnetic axes that differs between the MH and SH; the
results were interpreted either as nearly torque-free precession of a
slightly asymmetric crust \citep{Heyl2024} or as forced precession
of the crust \citep{Zhao2024}. \citet{Kolesnikov2022}, in turn,
reported 35-d-periodic pulse-frequency fluctuations in \fer/GBM data
consistent with triaxial free precession, and \textit{XRISM}
observations show a pulsation pattern of the Fe K line that evolves
with superorbital phase, which \citet{Kosec2026} interpret as the NS
accretion columns precessing alongside the disc in a ``double
precession'' framework.
On the theoretical side, \citet{Biryukov2026} showed that such free
precession requires the crustal superfluid to remain unpinned with an
anomalously low drag parameter ($\mathcal{R}\lesssim10^{-2}$) for
centuries, in tension with standard glitch physics, and
\citet{Markozov2026} found the associated continuous
gravitational-wave signal undetectable even for DECIGO-class
detectors. For the foreseeable future, therefore, the precession
hypothesis is testable only through X-ray timing and pulse-profile
studies such as this work, and its predicted signal vanishes
altogether if the clock is carried by the disc alone.

Independent support has just become available from the ELS itself:
the simultaneous IXPE/ART-XC/NuSTAR campaign during the present event
\citep{Tsygankov2026} found a reflection-dominated NuSTAR spectrum
and weak pulsations at the expected spin period in the ART-XC and
NuSTAR data: a blocked direct view with the disc clock still
running, exactly the configuration our monitoring results imply. The
measured period, $P = 1.237716(3)$~s at MJD~61221--61227, lies just
beyond the end of our GBM record and matches the extrapolated ELS
SD trend, confirming that the spin evolution continued
smoothly through the deepest part of the ELS.

Our monitoring results add three new, independent constraints to the
debate of accretion disc and NS precession.
First, the residual superorbital flux modulation stays
phase-locked to the pre-ELS clock as it fades through the 2025--2026
ELS (Sections~\ref{sec:els_ls} and \ref{sec:els_profile}):
the screen keeps its geometry and rhythm while the pulsar
fades from direct view, exactly what a precessing disc predicts,
with no need to invoke a change of the NS dynamics.
Second, at fixed superorbital phase the
pulse-profile morphology is statistically indistinguishable across
torque states and across the ELS transition
(Section~\ref{sec:profiles}): the beam pattern does not respond to
the accretion-rate collapse, so the flux decay is a screening effect
rather than a change of the emission geometry.
Third, and most directly relevant to the free-precession
interpretation, the profile shape shows no secular drift over the
18-yr \fer/GBM record (Section~\ref{sec:profiles}): the amplitude-based
parameters are stationary to within a fraction of a per cent per
decade, and the shift-invariant harmonic phases drift by less than a
degree per decade.
If the crust precessed with a geometry drifting at the rate implied by the IXPE
measurements \citep[$9.2\degr\pm1.8\degr$ between consecutive
years;][]{Heyl2024}, the pulse-profile morphology at fixed
superorbital phase should have evolved measurably over our much
longer baseline; the stability limits of Fig.~\ref{fig:stability}
constrain any such drift of the beam geometry to well below that
level.

Taken together, these results indicate that the precessing warped
disc alone suffices to account for the 35-d flux cycle, its
phase-locked fading through the ELS, and the unchanged pulse-profile
morphology, in line with the single-clock conclusion of
\citet{Staubert2013}. A freely precessing NS component is not
required by the present data. If it exists, its long-term geometric
drift must be small enough to leave both the pulse-profile shape and
the superorbital shape pattern unchanged at the level of our limits.
We note the limitations of our data set: it offers no polarization
channel, so the IXPE position-angle variations cannot be tested with
it. Moreover, because the phase zero of the public harmonic coefficients is
arbitrary from entry to entry, absolute pulse-arrival phases (the
most direct observable of a precessing beam) require
coherent-timing data from pointed instruments.

\subsection{What triggers the ELS?}
\label{sec:trigger}

If the fading clock of Sections~\ref{sec:els_ls} and
\ref{sec:els_profile} is carried by the disc, then the ELS is most
plausibly a transition of the disc's geometry.
The occulting structure is not an additional component, but the
same tilted, warped, precessing disc that produces the normal 35-d
cycle: its outer rim blocks the line of sight during the low states,
and its inner edge shapes the pulse-profile evolution. In the ELS, this
disc has migrated to a configuration that never clears our view.
That this picture suffices was already shown for the 1999--2000
event: the anomalous-low spectrum is the MH spectrum seen through
partial covering and heavy cold absorption, consistent with a
decreased disc inclination that obscures the central source over the
whole cycle \citep{Coburn2000}, while the reflection component
implies that the intrinsic luminosity barely changed
\citep{Still2001}.
Our results add the time dimension to this picture: the
visible MH decays smoothly to per-cent level within about eight superorbital cycles,
the fainter SH is buried first, the residual bump stays
phase-locked while it is detectable (Section~\ref{sec:els_ls}), and
the SD torque stays within its historical range
(Section~\ref{sec:nudot_flux}).
The ELS is therefore entered gradually,
over a few cycles, as a progressive closure of the direct
line of sight.

What pushes the disc over the edge? The historical record points at
the mass-transfer rate. Anomalous lows correlate with SD episodes and
with the minima of the turn-on O--C diagram, recurring on a
quasi-period of about five years \citep{Staubert2009}. The 1999--2000
event began with a rapid SD approaching the propeller regime
\citep{LeahyIgna2010}. The present event fits the same pattern and
sharpens it: the ELS coincides with the longest sustained SD,
$\dot{\nu} \approx -4.7 \times 10^{-13}$~Hz~s$^{-1}$ over
the $\gtrsim$380~d covered by the \fer/GBM record. The event itself has
lasted $\sim$480~d at the time of writing and is still ongoing.
The accretion flow reaching the magnetosphere thus genuinely decreased in
step with the flux collapse, but only moderately: the torque
stayed within the range of historical SD episodes while the flux fell
by an order of magnitude (Section~\ref{sec:nudot_flux}). The timing
decomposition is informative: the sustained SD begins already at the
cycle-212 turn-on, while the MH flux is still at 50--100 per cent of
normal, so the torque leads the flux collapse by a few cycles, as
expected if the accretion-rate change was established at the onset
and the screen closed progressively afterwards.
Within the irradiation-driven warping instability the warped disc is only
marginally stable \citep{Pringle1996,Wijers1999,Ogilvie2001}, and a
reduced stream impact both weakens the braking of the disc's
precession and lowers its tilt \citep{Klochkov2006}; the disc should
then precess faster, exactly as observed: the local cycle length at
the edge of the ELS, 34.564~d from the last 25 pre-ELS cycles
(Section~\ref{sec:els_profile}), is noticeably shorter than the
two-decade mean of 34.79~d, repeating the shortening inferred for the
same event in 1999--2000. A small perturbation of the mass-transfer
rate can thus be amplified into a full state switch through the
irradiation feedback loop between the disc and HZ Her.

Whether the torque decline drove the geometric closure, or both share
a common trigger, is not resolved by our data, which nevertheless
make a sharp, testable prediction for the recovery. If the disc merely
became opaque while keeping its clock, the ELS should end with an
abrupt turn-on close to the extrapolated phase, possibly carrying the
phase offset implied by the shortened period. The 1999--2000 event
indeed resumed after eighteen invisible cycles run at a shorter
period, with the disc clock having slipped by one cycle against the
pulse-profile clock \citep{Staubert2009}. A phase jump at the
recovery, or a long quiescent interval, would instead signal that
coherence was lost in the deep ELS state.

\section{Conclusions}
\label{sec:concl}

We have analysed two decades of \fer/GBM, \sw/BAT and \mx/GSC
monitoring of \hx, spanning the entry of the source into its first
ELS in 22 years. Our main results include:

\begin{enumerate}
	\item A significant $\sim$35-d superorbital modulation survives into
	the ELS: Lomb-Scargle periodograms detect it independently of any
	ephemeris. Cross-correlation timing further shows that the residual
	modulation stays phase-locked to the pre-ELS clock until its
	amplitude decays to per-cent level, when the in-phase coherence is
	lost (around cycles 219--221). The visible MH fades smoothly
	to per-cent level within about eight cycles of the onset, so the ELS
	is entered gradually, through a progressive closure of the direct
	line of sight.
	\item Across various torque states, the spin
	derivative correlates positively with the MH hard X-ray flux
	(Pearson $r\simeq0.7$ for both the \sw/BAT and \fer/GBM pulsed fluxes), the
	signature expected for disc accretion. The ELS extends the correlation
	into the low-flux regime without departing from it: the source shows
	its longest sustained SD of the record,
	$\dot{\nu}\approx-4.7\times10^{-13}$~Hz~s$^{-1}$ over the
	$\gtrsim$380~d of \fer/GBM coverage of the ELS, 	an event that has
	itself lasted $\sim$480~d at the time of writing. Its magnitude
	stays within the range of historical SD episodes.
	The sustained SD begins at the
	ELS onset while the MH flux is still essentially normal, so the
	accretion-rate change precedes the closure of the view by a few
	cycles.
	\item At fixed superorbital phase, the 8--50 keV pulse-profile
	morphology is statistically indistinguishable across the SU,
	SD, Flat and ELS samples, while the pulsed amplitude
	tracks the accretion state, and over the full 18-yr record the
	shape parameters are stationary to within a fraction of a per cent
	(amplitudes) and less than a degree (harmonic phases) per decade.
\end{enumerate}

Together, these results indicate that the precessing warped accretion
disc alone suffices to account for the 35-d flux cycle, its
phase-locked fading through the ELS, and the unchanged pulse-profile
morphology. The disc keeps its rhythm while the pulsar fades from
direct view, and the spin torque tracks the accretion flux through
the state transition.
A freely precessing NS component is not
required by the current monitoring data. If
it exists, its long-term geometric drift must
be small enough to leave the profile shape unchanged at the level of
our limits.

The eventual recovery from the present ELS will open a set of decisive
follow-up tests. (i) \emph{Clock memory}: turn-on timing and O--C
analysis of the first post-ELS cycles will show whether the 35-d clock
resumes at the phase extrapolated from the pre-ELS ephemeris, as it
did after the 1999--2000 event (which resumed after eighteen invisible
cycles at a shorter period with a one-cycle slip); a phase
jump would signal a realignment of the inner disc.
(ii) \emph{Torque recovery}: the SU that follows the recovery,
mapped against the MH flux, will populate the low-flux side of
the $\dot{\nu}$--$\mathcal{F}$ relation and test whether the decay
and recovery branches coincide. Any hysteresis between them would
mean that the screen's geometry depends on its own history: the
occulting screen would not reopen at the same accretion rate at which it
closed.
(iii) \emph{Geometry}: pulse-profile
and shape-parameter comparisons between the pre-ELS and recovered MH
will sharpen our stability limits, and \textit{IXPE} polarimetry of a
recovered MH can test directly whether the spin-axis geometry
resumed its pre-ELS configuration. With \sw\ temporarily out of
service, continued \mx/GSC and \fer/GBM monitoring, together with a prompt
pointed campaign at the first sign of a new turn-on, are the next steps.


\section*{Acknowledgements}

We acknowledge the support of the National Natural Science Foundation
of China (NSFC 12403051) and the Natural Science Foundation of Fujian
Province (2024J01902, 2026J0011109). During the preparation of this
work, the authors used Kimi K3 (Moonshot AI) for language polishing
of the manuscript and for assistance with the analysis code.
All AI-generated suggestions were reviewed and
verified by the authors against the original sources (including ADS
and the publishers' records), and the authors take full
responsibility for the content of this publication, including the
accuracy of all references.

\section*{Data Availability}

The data analysed in this work are publicly available from the following archives.
The \fer/GBM pulse-frequency and pulsed-flux history of \hx\ is provided by the GBM Accreting Pulsar Program\footnote{\url{https://gammaray.nsstc.nasa.gov/gbm/science/pulsars/lightcurves/herx1.html}}.
The \sw/BAT daily and orbit light curves are available through the BAT transient monitor\footnote{\url{https://swift.gsfc.nasa.gov/results/transients/HerX-1/}}.
The \mx/GSC daily and orbit light curves are publicly available from the MAXI web site\footnote{\url{http://maxi.riken.jp/star_data/J1657+353/J1657+353.html}}.
The derived data products, including the state classifications and the Lomb-Scargle periodogram results presented here, are available upon reasonable request to the corresponding author.



\bibliographystyle{mnras}
\bibliography{herX1_v3}


\clearpage
\bsp	
\label{lastpage}
\end{document}